\documentstyle[jkas]{article}

\runningauthor {H. KANG}
\runningtitle{DIFFUSIVE~SHOCK~ACCELERATION}

\month{October} \year{2012} \volume{45} \issueno{5}
\beginpage{111}\endpage{122}
\date{Received August 3, 2012; Revised September 12, 2012; Accepted September 13, 2012}

\def\kms{~{\rm km~s^{-1}}}
\def\cm3{~{\rm cm^{-3}}}
\def\yrs{~{\rm yr}}
\def\muG{~{\mu\rm G}}

\begin{document}

\title{DIFFUSIVE SHOCK ACCELERATION WITH MAGNETIC FIELD AMPLIFICATION AND ALFV\'ENIC DRIFT}

\author{Hyesung Kang}

\address{ Department of Earth Sciences, Pusan National University, Pusan  609
-735, Korea\\
 {\it E-mail : hskang@pusan.ac.kr}}

\address{\normalsize{\it (Received August 3, 2012; Revised  September 12, 2012; Accepted  September 13, 2012)}}

\abstract{
We explore how wave-particle interactions affect diffusive shock acceleration
(DSA) at astrophysical shocks 
by performing time-dependent kinetic simulations, 
in which phenomenological models for magnetic field amplification (MFA), Alfv\'enic drift, 
thermal leakage injection, Bohm-like diffusion, and a free escape boundary are implemented.
If the injection fraction of cosmic-ray (CR) particles is $\xi > 2\times10^{-4}$,
for the shock parameters relevant for young supernova remnants,
DSA is efficient enough to develop a significant shock precursor due to CR feedback,
and magnetic field can be amplified up to a factor of 20 via CR streaming instability
in the upstream region.
If scattering centers drift with Alfv\'en speed
in the amplified magnetic field, the CR energy spectrum can be steepened significantly
and the acceleration efficiency is reduced.
Nonlinear DSA with self-consistent MFA and Alfv\'enic drift
predicts that the postshock CR pressure saturates roughly at $\sim 10$ \% of the shock ram pressure
for strong shocks with a sonic Mach number ranging $20\la M_s\la 100$. 
Since the amplified magnetic field follows the flow modification in the precursor,
the low energy end of the particle spectrum is softened much more than the high energy end. 
As a result, the concave curvature in the energy spectra does not disappear entirely
even with the help of Alfv\'enic drift.
For shocks with a moderate Alfv\'en Mach number ($M_A<10$), 
the accelerated CR spectrum can become as steep as $E^{-2.1}-E^{-2.3}$,
which is more consistent with the observed CR spectrum and gamma-ray photon spectrum of
several young supernova remnants.
}

\maketitle

\section{INTRODUCTION}

Diffusive shock acceleration (DSA) theory explains how nonthermal particles are produced 
through their interactions with MHD waves in the converging flows 
across collisionless shocks in astrophysical plasmas \citep{bell78, dru83, blaeic87}.
Theoretical studies have shown that 
some suprathermal particles with velocities large enough to swim against the
downstream flow can return across the shock and stream upstream, 
and that streaming motions of high energy particles 
against the background fluid generate both resonant and nonresonant waves 
upstream of the shock \citep{bell78,lucek00, bell04, riqu09,roga12}.
Those waves in turn scatter CR particles and amplify turbulent magnetic fields in the preshock region. 
These plasma physical processes, i.e., injection of suprathermal particles into the 
CR population, self-excitation of MHD waves, and amplification of magnetic fields
are all integral parts of DSA \citep[e.g.,][]{maldru01}.

Multi-band observations of nonthermal radio to $\gamma$-ray
emissions from supernova remnant (SNR) shocks have confirmed the acceleration
of CR electrons and protons up to $\sim100$ TeV \citep[e.g.,][]{abdo10,abdo11,acero10,
acciari11,giordano12}.
Moreover, thin rims of several young SNRs in high-resolution X-ray observations
indicate the presence of downstream magnetic fields as strong as a few $100 \mu$G, 
implying efficient magnetic field amplification (MFA) at these shocks \citep[e.g.,][]{parizot06,eriksen11,reynolds12}.

The most attractive feature of the DSA theory is the simple prediction of 
power-law energy spectra of CRs, $N(E) \propto E^{-(\sigma+2)/(\sigma-1)}$
(where $\sigma$ is the shock compression ratio) in the
test particle limit. For strong, adiabatic gas shocks with $\sigma=4$, this gives a power-law
index of 2, which is reasonably close to the observed `universal' index
of the CR spectra in many environments.
However, nonlinear treatments of DSA predict that
at strong shocks there are highly nonlinear back-reactions from CRs to the underlying flow,
creating a shock precursor \citep[e.g.,][]{bv97,kj07}.
So the particles just above the injection momentum ($p_{\rm inj}$)
sample mostly the compression across the subshock ($\sigma_s$),
while those near the highest momentum ($p_{\rm max}$)
experience the greater, total compression across the entire shock structure
($\sigma_t$).
This leads to the CR energy spectrum that
behaves as $N(E)\propto E^{-(\sigma_s+2)/(\sigma_s-1)}$ for $p\sim p_{\rm inj}$,
but flattens gradually to $ N(E)\propto E^{-(\sigma_t+2)/(\sigma_t-1)}$ toward
$p\sim p_{\rm max}$ \citep{kang09}.
For example, the power-law index becomes 1.5 for $\sigma_t=7$.

In contrast to such expectations, however, the GeV-TeV $\gamma$-ray spectra of several 
young SNRs seem to require the proton spectrum 
as steep as $N(E)\propto E^{-2.3}$, if the observed $\gamma$-ray
photons indeed originate from $\pi^0$ decay \citep{abdo10,giordano12}.
This is even softer than the test-particle power-law for strong shocks.
Moreover, \citet{ave09} showed that the spectrum of CR nuclei
observed at Earth can be fitted by a single power law of $J(E) \propto E^{-2.67}$ below $10^{14}$ eV.
Assuming an energy-dependent propagation path length ($\Lambda \propto E^{-0.6}$), they
suggested that a soft source spectrum, $N(E)\propto E^{-\alpha}$ with $\alpha
\sim 2.3-2.4$ is preferred by the observed data.
These observational data appear to be inconsistent with flat CR spectra predicted 
by nonlinear DSA model for the SNR origin of Galactic CRs.

It has been suggested that non-linear wave damping and wave dissipation due to ion-neutral 
collisions may weaken the stochastic scattering on relevant scales, 
leading to slower acceleration than predicted based on the the so-called Bohm-like diffusion,
and escape of the highest energy particles from the shock \citep[e.g.][]{pz05,capri09}. 
These processes may lead to the particle energy spectrum at the highest energy end that is
much steeper than predicted by nonlinear DSA.
Escape of high energy protons from SNRs is an important yet very complex problem 
that needs further investigation \citep{malkov11, dru11}.

Recently some serious efforts have been underway 
to understand at least some of the complex plasma processes through Particle-in-Cell (PIC) and hybrid plasma 
simulations \citep[e.g.][]{riqu09,guo10,garat12}. 
However, these types of plasma simulations are too much demanding and too
expensive to study the full extent of the DSA problem. So we do not yet understand them in enough 
detail to make precise quantitative predictions for the injection and acceleration rate and efficiency.
Instead, most of kinetic approaches commonly adopt 
phenomenological models that can emulate more or less self-consistently
some of those plasma interactions, for example, 
the thermal leakage injection, magnetic field amplification, wave-damping and etc
\citep[e.g.,][]{bkv09,kang10,pzs10,lee12,capri12}.   

In our previous studies, we considered DSA of CR protons, assuming that magnetic field strength is
uniform in space and constant in time without self-consistent MFA \citep[e.g.][]{kj07, kang09}.
In the present paper, we explore how the following processes
affect the energy spectra of CR protons and electrons accelerated at plane astrophysical shocks: 
1) magnetic field amplified by CR streaming instability in the precursor 
2) drift of scattering centers with Alfv\'en speed in the {\it amplified magnetic field}, 
and 3) escape of highest energy particles from the shock. 
Toward this end we have performed time-dependent numerical simulations, in which
DSA of CR protons and electrons at strong planar shocks 
is followed along with electronic synchrotron and inverse Compton (IC) losses. 
Magnetic field amplification due to resonant waves generated by CR streaming
instability is included through an approximate, analytic model suggested by
\citet{capri12}. Escape of highest energy particles near maximum momentum, $p_{\rm max}$,
is included by implementing a free escape boundary (FEB) at a upstream location. 
As in our previous works \citep[e.g.][]{kang10,kang11}, a thermal leakage injection model, 
a Bohm-like diffusion coefficient ($\kappa(p) \propto p$), 
and a model for wave dissipation and heating of the gas are adopted as well.

In the next section we describe the numerical method and 
phenomenological models for the plasma interactions in DSA theory,
and the model parameters for planar shocks.  
The simulation results will be discussed in Section 3, 
followed by a brief summary in Section 4.

\section{DSA MODEL}
\subsection{CRASH Code for DSA}

Here we consider the CR acceleration at quasi-parallel shocks
where the mean background magnetic field lines are parallel to the shock normal.
So we solve the standard gasdynamic equations with CR proton pressure terms
added in the conservative, Eulerian formulation for one dimensional plane-parallel geometry.
The basic gasdynamic equations and details of the CRASH (Cosmic-Ray Amr SHock) code
can be found in \citet{kjg02} and \citet{kang11}. 

We solve the following diffusion-convection equations
for the pitch-angle-averaged phase space distribution function
for CR protons, $g_p=f_p p^4$, and for CR electron, $g_e=f_e p^4$
\citep{skill75}:

\begin{eqnarray}
{\partial g\over \partial t}  + (u+u_w) {\partial g \over \partial x}
= {1\over{3}} {\partial \over \partial x} (u+u_w) \left( {\partial g\over
\partial y} -4g \right) \nonumber\\
 + {\partial \over \partial x} \left[\kappa(x,y)
{\partial g \over \partial x}\right]
+ p {\partial \over {\partial y}} \left( {b\over p^2} g \right) ,
\label{dc}
\end{eqnarray}
where $y=\ln(p/m_p c)$.
Here the particle momentum is expressed in units of $m_pc$
and so the spatial diffusion coefficient, $\kappa(x,p)$, has the same form
for {\it both protons and electrons}.
The velocity $u_w$ represents the effective relative motion of
scattering centers with respect to the bulk flow velocity, $u$,
which will be described in detail in section 2.5.

The cooling term $b(p)=-dp/dt$ takes account for electron
synchrotron/IC losses, while it is set to be $b(p)=0$ for protons.
Here the synchrotron/IC cooling constant for electrons is defined as
\begin{equation}
b(p) = \frac{4 e^4 }{9 m_e^4 c^6} B_{\rm e}^2 p^2
\label{ecool}
\end{equation}
in cgs units, where $e$ and $m_e$ are electron charge and mass, respectively.
Here $B_{\rm e}^2= B^2 + B_r^2$ as the effective magnetic field strength 
for radiative losses
including the energy density of the ambient radiation field.
We set $B_r=6.5\muG$, including the cosmic background and mean Galactic
radiation fields \citep{ekjm11}.

The dynamical effects of the CR proton pressure are included
in the DSA simulations,
while the CR electrons are treated as test-particles.
In order to include the dynamical effects of amplified magnetic field, the magnetic
pressure, $P_B=B^2/8\pi$, is added to the momentum conservation equation as follows:
\begin{equation}
{\partial (\rho u) \over \partial t}  +  {\partial (\rho u^2 + P_g + P_c + P_B) \over \partial x} = 0.
\label{mocon}
\end{equation}
However, our model magnetic field amplification typically results in
$P_{B}/\rho_0 u_s^2 < 0.01$ in the precursor, where $\rho_0 u_s^2$ is the shock ram pressure (see Section 2.4).

\subsection{Thermal Leakage Injection}

The injection rate with which suprathermal particles are injected into CRs
at the subshock depends in general upon the shock Mach number,
field obliquity angle, and strength of Alfv\'enic turbulence responsible
for scattering.
In thermal leakage injection models
suprathermal particles well into the exponential tail of the postshock Maxwellian
distribution leak upstream across a quasi-parallel shock \citep{maldru01,kjg02}.
We adopt a simple injection scheme in which the particles above an
effective injection momentum $p_{\rm inj}$ cross the shock and get
injected to the CR population:
\begin{equation}
p_{\rm inj} \approx 1.17 m_p u_2 \left(1+ {1.07 \over \epsilon_B} \right),
\label{pinj}
\end{equation}
where the injection parameter, $\epsilon_B = B_0/B_{\perp}$, is the ratio of
the large-scale magnetic field along the shock normal, $B_0$, to
the amplitude of the postshock MHD wave turbulence, $B_{\perp}$ \citep{kjg02}.
With a larger value of $\epsilon_B$ (i.e., weaker turbulence),
$p_{\rm inj}$ is smaller, which results in a higher injection rate.
We consider $\epsilon_B=0.23$ here.

We define the injection efficiency as the fraction of particles that have
entered the shock from far upstream and then injected into the CR distribution:
\begin{equation}
\xi(t)=\frac {\int {d x} \int_{p_{\rm inj}}^{\infty} 4\pi f_p(p,x,t)
p^2 {d p}}
 {n_0 u_s t }\,
\label{xieq}
\end{equation}
where $n_0$ is the particle number density far upstream and $u_s$ is the shock speed.

Since postshock thermal electrons need to be pre-accelerated
before they can be injected into Fermi process, 
it is expected that electrons are
injected at the shock with a much smaller injection rate, i.e., 
the CR electron-to-proton ratio is estimated to be small, 
$K_{e/p} \sim 10^{-4}-10^{-2}$ \citep{reynolds08,morlino12}.
Since this ratio is not yet constrained accurately by plasma physics
and we do not consider nonthermal emissions from CR particles in this paper,
both protons and electrons are injected in the same manner in our simulations 
(i.e. basically $K_{e/p}=1$).
But $K_{e/p}=0.1$ will be used just for clarity of some figures below. 

\subsection{Bohm-like Diffusion Model}

It is assumed that CR particles are resonantly scattered by Alfv\'en waves,  
which are excited by CR streaming instability in the upstream region 
and then advected and compressed in the down stream region
\citep{bell78,lucek00}.
So in DSA modeling the Bohm diffusion model, $\kappa_B= (1/3) r_g v$, is commonly used to represent 
a saturated wave spectrum.
We adopt a Bohm-like diffusion coefficient
that includes a flattened non-relativistic momentum dependence,
\begin{equation}
\kappa(x,p) = \kappa_{\rm n} {B_0 \over B(x)} \cdot {p \over m_pc}, 
\label{Bohm}
\end{equation}
where $\kappa_n= m_p c^3/(3eB_0)= (3.13\times 10^{22} {\rm cm^2s^{-1}}) B_0^{-1}$,
and $B_0$ is the magnetic field strength far upstream expressed in units of 
microgauss.
The local magnetic field strength, $B(x)$, will be described in the next
section. 
Hereafter we use the subscripts `0', `1', and `2' to denote
conditions far upstream of the shock, immediately upstream and downstream of the subshock, respectively.

\subsection{Magnetic Field Amplification}

Since the resonant interactions amplify mainly the turbulent magnetic field perpendicular to 
the shock normal in the quasi-linear limit, it was commonly assumed that the parallel component is $B_{\parallel} 
\approx B_0$, 
the unperturbed upstream field \citep{capri09}.
In a strong MFA case, however, the wave-particle interaction and the CR transport are not 
yet understood fully.
For example, plasma simulations by \citet{riqu09} showed that both $B_{\parallel}/B_0$
and $B_{\perp}/B_0$ can increase to $\sim 10-30$ via Bell's CR current driven instability.
Here we follow the prescription for MFA that was formulated by \citet{capri12}
based on the assumption of isotropization of amplified magnetic field.

In the upstream region ($x>x_s$),
\begin{equation}
{B(x)^2 \over B_0^2} = 1 + (1-\omega_H)\cdot {4\over 25}M_{A,0}^2 { {(1-U(x)^{5/4})^2}\over U(x)^{3/2}},
\label{Bpre}
\end{equation}
where $M_{A,0}= u_s/v_{A,0}$ is the Alfv\'en Mach number for the far upstream Alfv\'en speed, 
$v_{A,0}= B_0/ \sqrt{4\pi \rho_0}$,
and $U(x)= [u_s-u(x)]/u_s$ is the flow speed in the shock rest frame normalized by
the shock speed. 
The factor $(1-\omega_H)$ is introduced to take account of the loss of magnetic energy due to wave dissipation,
which will be discussed in Section 2.5. 
Obviously, $\omega_H=0$ means no dissipation, while $\omega_H=1$ means
complete dissipation of waves (i.e., no MFA). 
Here $\omega_H= 0.1$ will be considered as a fiducial case, since we are interested in the case
where the effects of MFA and ensuing wave drift are the greatest.

This MFA model predicts no amplification in the test-particle regime, where the flow structure 
is not modified (i.e., $U(x)=1$). 
In the case of ``moderately modified'' shocks,  for example, if $U_1\approx 0.8$ and $\omega_H=0.1$, 
the amplified magnetic field strength scales as $B_1/B_0 \approx 0.11 M_{A,0}$.
So for $M_{A,0}\approx 150$, the preshock amplification factor could become $B_1/B_0 \approx 17$.
On the other hand, the ratio of the magnetic pressure to the shock ram pressure becomes 
$P_{B,1}/\rho_0 u_s^2 = (2/25)(1-U_1^{5/4})^2/U_1^{3/2}\approx 6.6\times 10^{-3}$.
So we expect that even the amplified field is not dynamically important in the precursor.

The magnetic filed strength immediately upstream of the subshock, $B_1$, is estimated
by Equation (\ref{Bpre}) and assumed to be completely turbulent.
Moreover, assuming that the two perpendicular components are simply compressed 
across the subshock, 
the immediate postshock field strength can be estimated by
\begin{equation}
B_2/B_1=\sqrt{1/3+2/3(\rho_2/\rho_1)^2}.
\label{B2}
\end{equation}
So for the case with $\rho_2/\rho_1\approx 4.2$, $B_2/B_1\approx 3.5$.
Then we assume in the downstream region the field strength scales with the gas density:   
\begin{equation}
B(x) = B_2 \cdot \left[\rho(x)/ \rho_2 \right].  
\label{Bpost}
\end{equation}

We note the MFA model described in Equations (\ref{Bpre})-(\ref{Bpost}) is used also
for the diffusion coefficient model given in Equation (\ref{Bohm}). 
Hence the maximum momentum $p_{\rm max}$ is controlled by the degree of MFA as well.

\begin{table*}
\begin{center}
{\bf Table 1.}~~Model Parameters$^{\rm a}$\\
\vskip 0.3cm
\begin{tabular}{ lrrrrrrrrr }
\hline\hline
Model$^{\rm b}$ & $u_s$ & $n_H$ (ISM) & $T_0$ & $M_s$ & $M_{A,0}$ & 
 $f_A~ ^{\rm c}$ & ~$\omega_H~ ^{\rm d}$ & $u_{w,1}$ & $u_{w,2}$ \\
~& $\kms$ & $(\rm cm^{-3})$ & (K) &
 ~ & ~ &  &  & & \\

\hline
W1a  &$3\times10^3$  & 1.0  &$4.0\times10^4$  &100 & 164. & 1.0 & 0.1  & $+v_A$ &  $0$  \\
W1b  &$3\times10^3$  & 1.0  &$4.0\times10^4$ &100 & 164. & 1.0 & 0.1  & $0$ &  $0$  \\
H1a &$3\times10^3$ & 1.0   & $10^6$ &20 &  164. & 1.0 & 0.1 &  $+v_A$ &  $0$  \\
H1b &$3\times10^3$ & 1.0   & $10^6$ &20 &  164. & 1.0 & 0.1 &  $0$   &  $0$   \\
H1c &$3\times10^3$ & 1.0   & $10^6$ &20 &  164. & 0.5 & 0.5 &  $+v_A$ &  $0$ \\
H2a &$3\times10^3$ & 0.01  & $10^6$ &20 &  16.4 & 1.0 & 0.1 &  $+v_A$ &  $0$ \\
H2b &$3\times10^3$ & 0.01  & $10^6$ &20 & 16.4 & 1.0 & 0.1 &  $0$    &  $0$ \\
H2d &$3\times10^3$ & 0.01  & $10^6$ &20 & 16.4 & 1.0 & 0.1 &  $+v_A$  & $-v_A$ \\
H3a &$10^3$ & 0.01   & $10^6$ &6.67 & 5.46 & 1.0 & 0.1 &  $+v_A$ &  $0$ \\
H3b &$10^3$ & 0.01   & $10^6$ &6.67 & 5.46 &  1.0 & 0.1 &  $0$  &  $0$\\
H4a &$4.5\times10^3$ & 0.01  & $10^6$ &30 &  24.6 & 1.0 & 0.1 &  $+v_A$ &  $0$ \\
H4b &$4.5\times10^3$ & 0.01  & $10^6$  &30 & 24.6 & 1.0 & 0.1 &  $0$    &  $0$\\

\hline
\end{tabular}
\end{center}
$^{\rm a}$ For all the models the background magnetic field is $B_0=5\muG$ and the injection parameter is $\epsilon_B=0.23$.\\ 
$^{\rm b}$ `W' and `H' stands for the warm and hot phase of the ISM, respectively.\\ 
$^{\rm c}$ See  (9) for the Alfv\'en parameter.\\
$^{\rm d}$ See Equation (6) for the wave dissipation parameter. \\ 

\end{table*}

\begin{figure*}[t]
\vskip -1.0cm
\centerline{\epsfysize=14cm\epsfbox{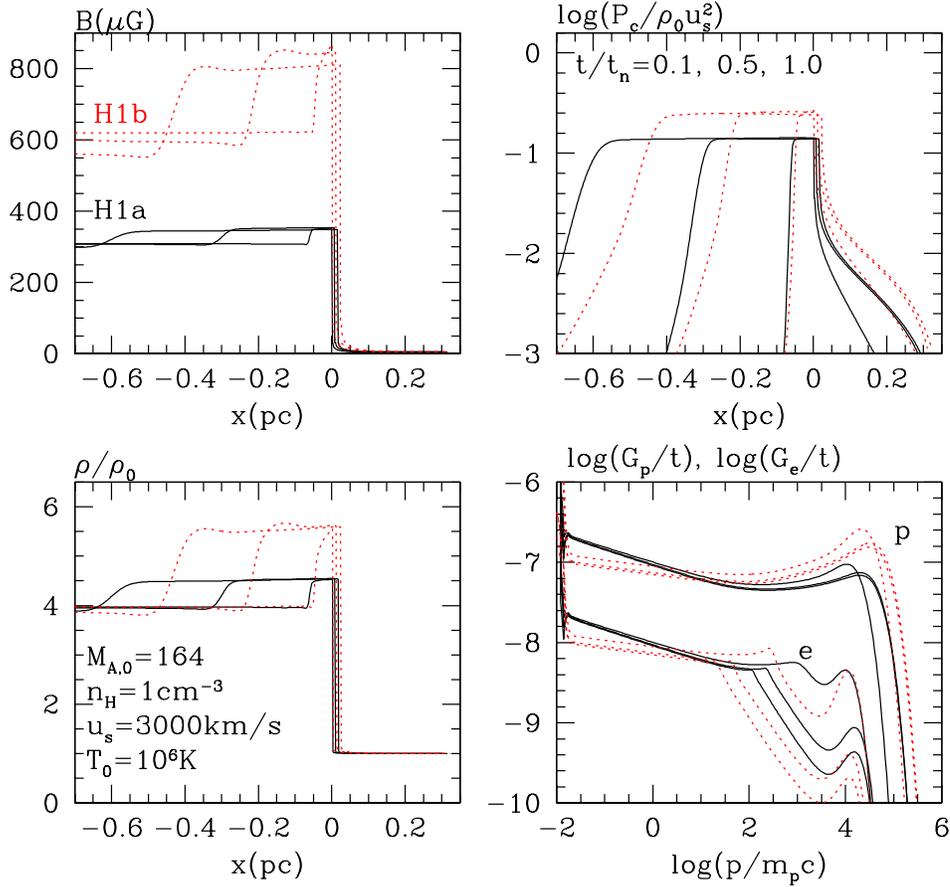}}
\vskip -1.5cm
\caption{
Time evolution of the magnetic field strength, CR pressure, gas density, and volume integrated distribution 
functions of protons ($G_p$) and electrons ($G_e$) for H1a (solid lines) and H1b (dotted lines) models at $t/t_n =  ~0.1,~2.5$ and 5.
See Table 1 for other model parameters and normalization constants.
In the bottom right panel the upper curves are for the proton spectra, while the lower curves are for the electron
spectra. 
Note that both $G_p/t$ and $G_e/t$ are given in arbitrary units and
$K_{e/p}=0.1$ is adopted here for clarity.
}
\label{fig1}
\end{figure*}
\begin{figure*}[t]
\vskip -1.0cm
\centerline{\epsfysize=14cm\epsfbox{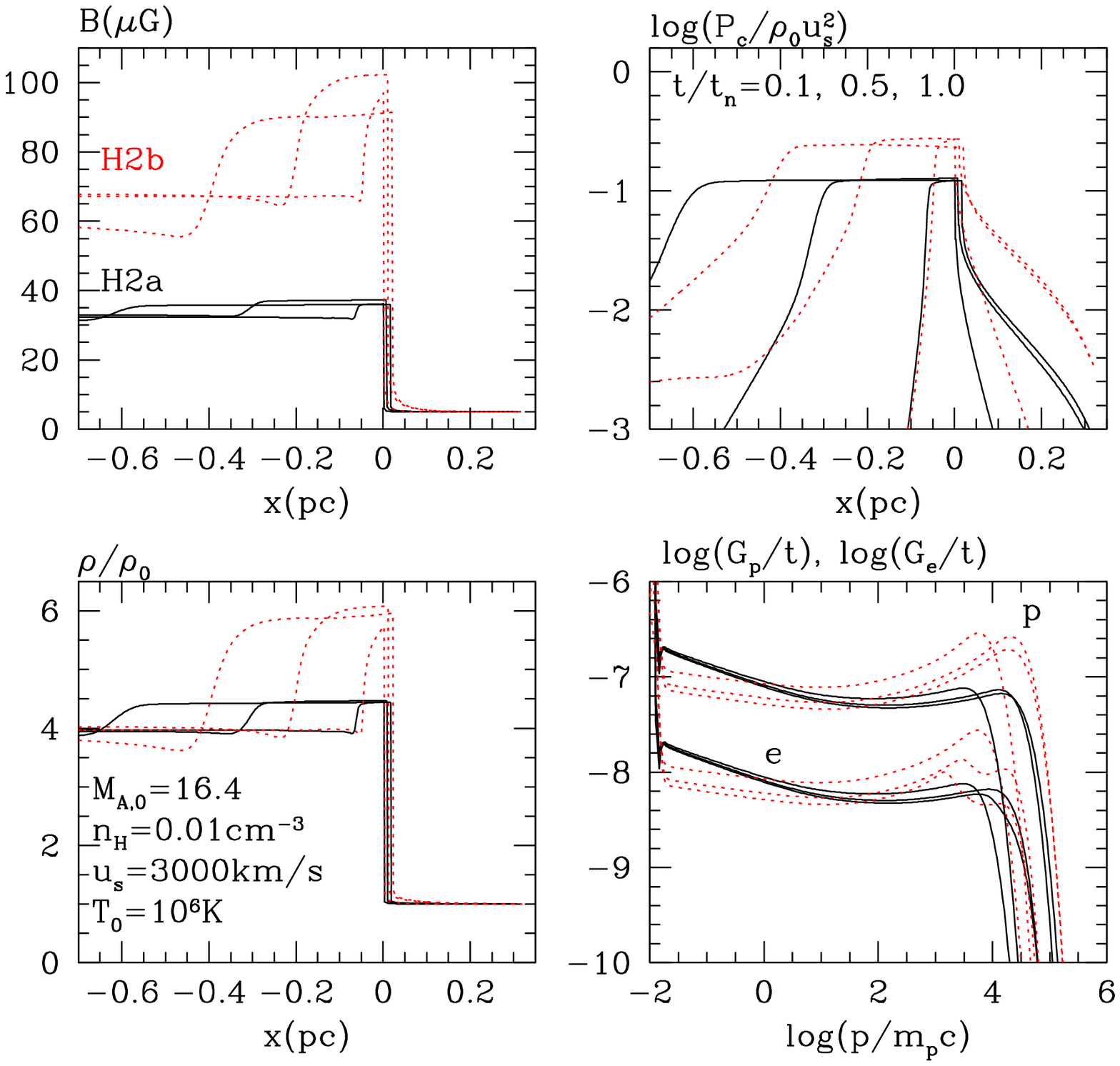}}
\vskip -1.5cm
\caption{
Same as Figure 1 except that H2a (solid lines) and H2b (dotted lines) are shown.
}
\label{fig2}
\end{figure*}
\begin{figure*}[t]
\vskip -1.0cm
\centerline{\epsfysize=14cm\epsfbox{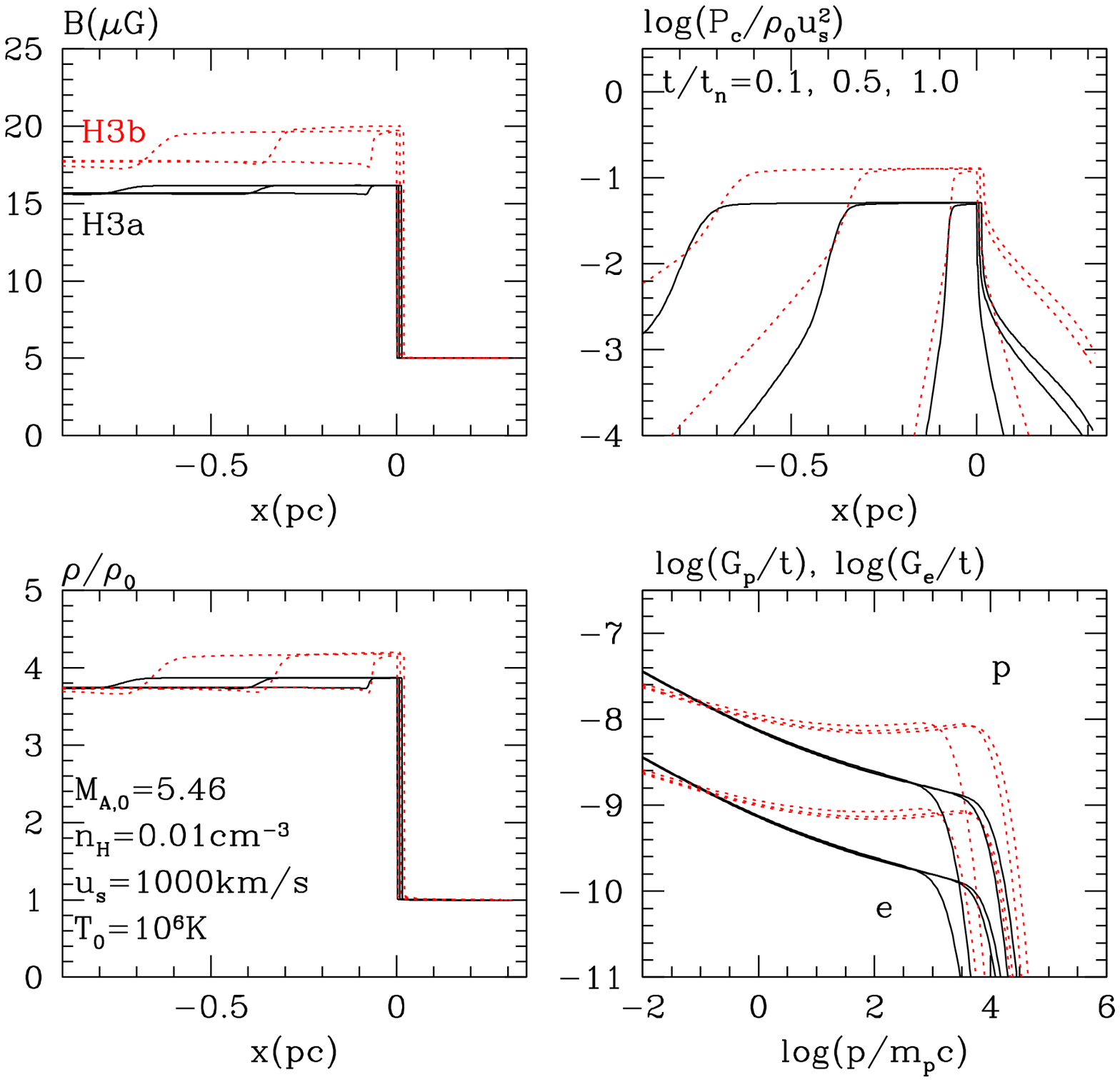}}
\vskip -1.5cm
\caption{
Same as Figure 1 except that H3a (solid lines) and H3b (dotted lines) are shown.
}
\label{fig3}
\end{figure*}

\subsection{Alfv\'enic Drift}

The resonant waves generated by CR streaming instability will drift with
respect to the underlying flow and also transfer energy
to the gas through dissipation \citep[e.g.][]{skill75,jon93}.
These two effects influence the accelerated particle spectrum and the DSA efficiency as follows. 
The scattering by Alfv\'en waves tends to isotropize
the CR distribution in the wave frame rather than in the gas frame \citep{bell78},
which reduces the velocity difference between upstream and downstream scattering centers,
compared to that of the bulk flow.
The resulting CR spectrum becomes softer than estimated without considering
the wave drift. 

The mean drift speed of scattering centers is commonly set to be the Alfv\'en speed, i.e.,
$u_{w,1}(x) = + v_A = + B_{\parallel} / \sqrt{4\pi \rho}$, pointing away from the shock, 
where $B_{\parallel}$ is the local magnetic field strength parallel to the shock normal.
As described in Equation (\ref{Bpre}) here we assume that both $B_{\parallel}$
and $B_{\perp}$ are amplified and isotropized, so scattering centers drift with Alfv\'en speed
in the local amplified magnetic field.
In order to take account of the uncertainty regarding this issue, we model
the local Aflv\'en speed as
\begin{equation}
 v_A(x) = { {B_0 + (B(x)-B_0)f_A} \over \sqrt{4\pi \rho(x)}},
\label{vA}
\end{equation}
where the parameter $f_A$ is a free parameter
\citep{zp08,lee12}.
If scattering centers drift along the amplified field ($f_A=1$), the Alfv\'enic drift 
will have the maximum effects.
Here we will consider the models with $f_A=0.5-1.0$ (see Table 1). 

In the postshock region the Alfv\'enic turbulence is probably relatively balanced, 
so the wave drift can be ignored, that is, $u_{w,2} \approx 0$ \citep{jon93}.
On the other hand, if the scattering centers drift away from the shock in both upstream and downstream regions,
the accelerated particle spectrum could be softened drastically \citep[e.g.][]{zp08}.
We will consider one model (H2d) in which $u_{w,2} \approx - v_A$ is adopted in the
downstream of the shock (see Table 1).

As mentioned in the Introduction, the CR spectrum develops a concave curvature when the preshock flow is modified
by the CR pressure.
If we include the Alfv\'enic drift only in the upstream flow, the slope of the momentum distribution function,
$q=-\partial \ln f/\partial \ln p$, can be estimated as
\begin{equation}
q_s \approx {{3(u_1-u_{w,1})}\over (u_1-u_{w,1})-u_2}
\approx {{3\sigma_s (1-M_{A,1}^{-1})}\over \sigma_s(1-M_{A,1}^{-1})-1 }
\label{qs}
\end{equation}
for $p\sim p_{\rm inj}$, and
\begin{equation}
q_t \approx {{3(u_0-u_{w,0})}\over (u_0-u_{w,0})-u_2}
\approx {{3\sigma_t (1-M_{A,0}^{-1})}\over \sigma_t(1-M_{A,0}^{-1})-1}
\label{qt}
\end{equation}
for $p \sim p_{\rm max}$.
Here $M_{A,1}= u_1/v_{A,1}$ is Alfv\'enic Mach number immediately upstream of the subshock.
As can be seen in these equations, a significant steepening will occur only if $M_A \la 10$ 
\citep{capri12}. 

According to the MFA prescription given in Equation (\ref{Bpre}), 
the amplification factor depends on the precursor modification, that is,
the ratio $B(x)/B_0$ is unity far upstream and increases through the precursor toward the subshock. 
So the Afv\'enic drift speed is highest immediately upstream of
the subshock, while it is the same as the unperturbed Alfv\'en speed, $v_{A,0}$ at the far upstream region
($M_{A,1}\ll M_{A,0}$).
Thus the Alfv\'enic drift is expected to steepen preferentially the lower energy end of the CR spectrum, 
since the lowest energy particles diffuse mostly near the subshock. 
For the highest energy particles, which diffuse
over the distance of $\sim \kappa(p_{\rm p,max})/u_s$, however, 
the Alf\'evnic drift does not steepenthe CR spectrum significantly,
if $M_{A,0}\gg 1$.

\begin{figure*}[t]
\vskip -1.0cm
\centerline{\epsfysize=15cm\epsfbox{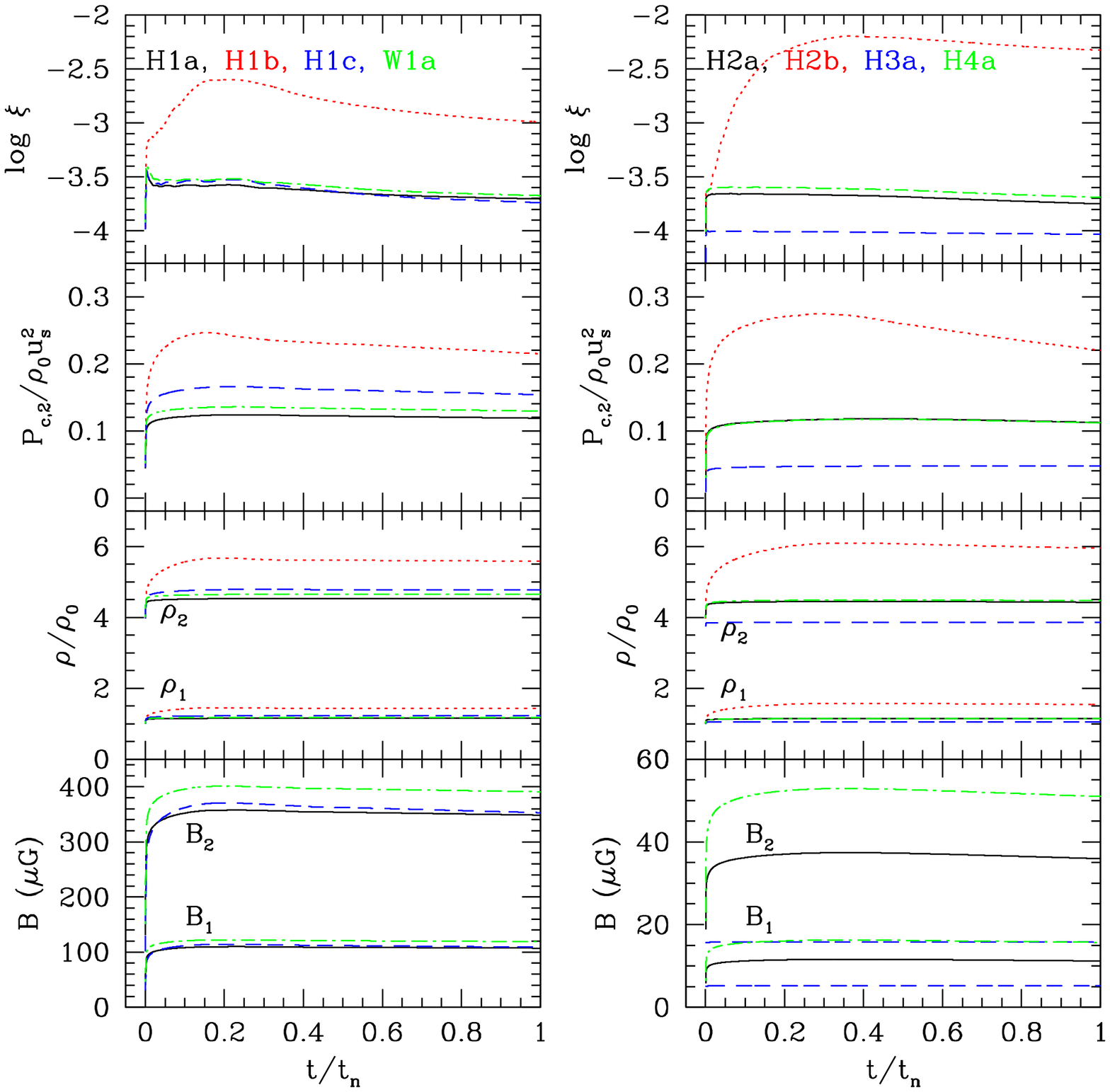}}
\vskip -0.3cm
\caption{
Time evolution of the injection efficiency, $\xi$, postshock CR pressure, $P_{c,2}$,
the gas density $\rho_1$ ($\rho_2$) immediately upstream (downstream) of the subshock,
and magnetic field strengths, $B_1$ and $B_2$, in different models:
H1a (black solid lines), H1b (red dotted), H1c (blue dashed), W1a (green dot-dashed) in the left column,
and H2a (black solid lines), H2b (red dotted), H3a (blue dashed), H4a (green dot-dashed) in the right column.
Note H1b and H2b models are shown for comparison, but $B_1$ and $B_2$ for those models are not included in the bottom panels.
}
\label{fig4}
\end{figure*}

\subsection{Wave Dissipation and Particle Escape}

As discussed in the Introduction, non-linear wave damping and dissipation 
due to ion-neutral collisions may weaken the stochastic scattering, 
leading to slower acceleration and escape of highest energy particles from the shock.
These processes are not understood quantitatively well, 
so we adopt a simple model in which waves are dissipated locally as heat in the precursor. 
Then gas heating term in the upstream region is prescribed as 
\begin{equation}
W(x,t)= -  \omega_H \cdot v_A(x) {\partial P_c \over \partial x },
\end{equation}
where $P_c$ is the CR pressure \citep{jon93}.
The parameter $\omega_H$ is introduced to control the degree of wave dissipation
and a fiducial value of $\omega_H=0.1$ is assumed.
As shown previously in SNR simulations \citep[e.g.][]{bv97,kj06}, 
this precursor heating reduces the subshock Mach number thereby 
reducing the DSA efficiency.
For larger values of $\omega_H$, the magnetic field amplification is suppressed (see Equation (\ref{Bpre})), which also
reduces the maximum momentum of protons and so the DSA efficiency.

In addition, we implement a free escape boundary (FEB) at a upstream location
by setting $f(x_{\rm FEB},p)=0$ at $x_{\rm FEB}= 0.1 R_s=0.3 {\rm pc}$ (here the shock is located at $x_s=0$).
This FEB condition can mimic the escape of the highest energy particles with the diffusion length, 
$\kappa(p)/u_s \ga x_{\rm FEB}$.
For typical supernova remnant shocks, this FEB leads to the size-limited maximum momentum, 
\begin{equation}
{p_{\rm p,max}\over m_p c} \approx 4.4\times 10^4 ({B_0\over 5\mu{\rm G}})({u_s\over 3000 \kms})({x_{\rm FEB}\over 0.3{\rm pc}}).
\end{equation}
As can be seen in Section 3, the CR proton spectrum and the shock structure approach to time-asymptotic states, if
this FEB is employed \citep{kang09}.

On the other hand, the maximum electron momentum can be estimated by 
\begin{equation}
{p_{\rm e,max} \over {m_pc}} \approx 2.8\times 10^4
 \left({B_1 \over 30 \muG}\right)^{-1/2} \left({u_s \over {3000 \kms}}\right),
\label{peq}
\end{equation}
which is derived from the equilibrium condition that the DSA momentum gains per cycle
are equal to the synchrotron/IC losses per cycle \citep{kang11}.
The electron spectrum {\it at the shock position}, $f_e(x_s,p)$, cuts off exponentially at $\sim p_{\rm e,max}$.

On the other hand, the postshock electron spectrum cuts off at a progressively lower momentum
downstream from the shock due to the energy losses.
That results in the
steepening of the volume integrated electron energy spectrum, $F_e(p)=\int f_e(x,p) dx $, by one power of the momentum \citep{kang12}.
At the shock age $t$, the break momentum can be estimated from the condition $t=p/b(p)$:
\begin{equation}
{p_{\rm e,br}(t) \over {m_pc}} \approx 1.3 \times 10^3 \left({t \over 10^3 \yrs}\right)^{-1} \left({B_{\rm e,2} \over
{100 \muG}}\right)^{-2},
\label{pbr}
\end{equation}
which depends only on the postshock magnetic field strength and the shock age \citep{kang11}.

\begin{figure*}[t]
\vskip -1.0cm
\centerline{\epsfysize=14cm\epsfbox{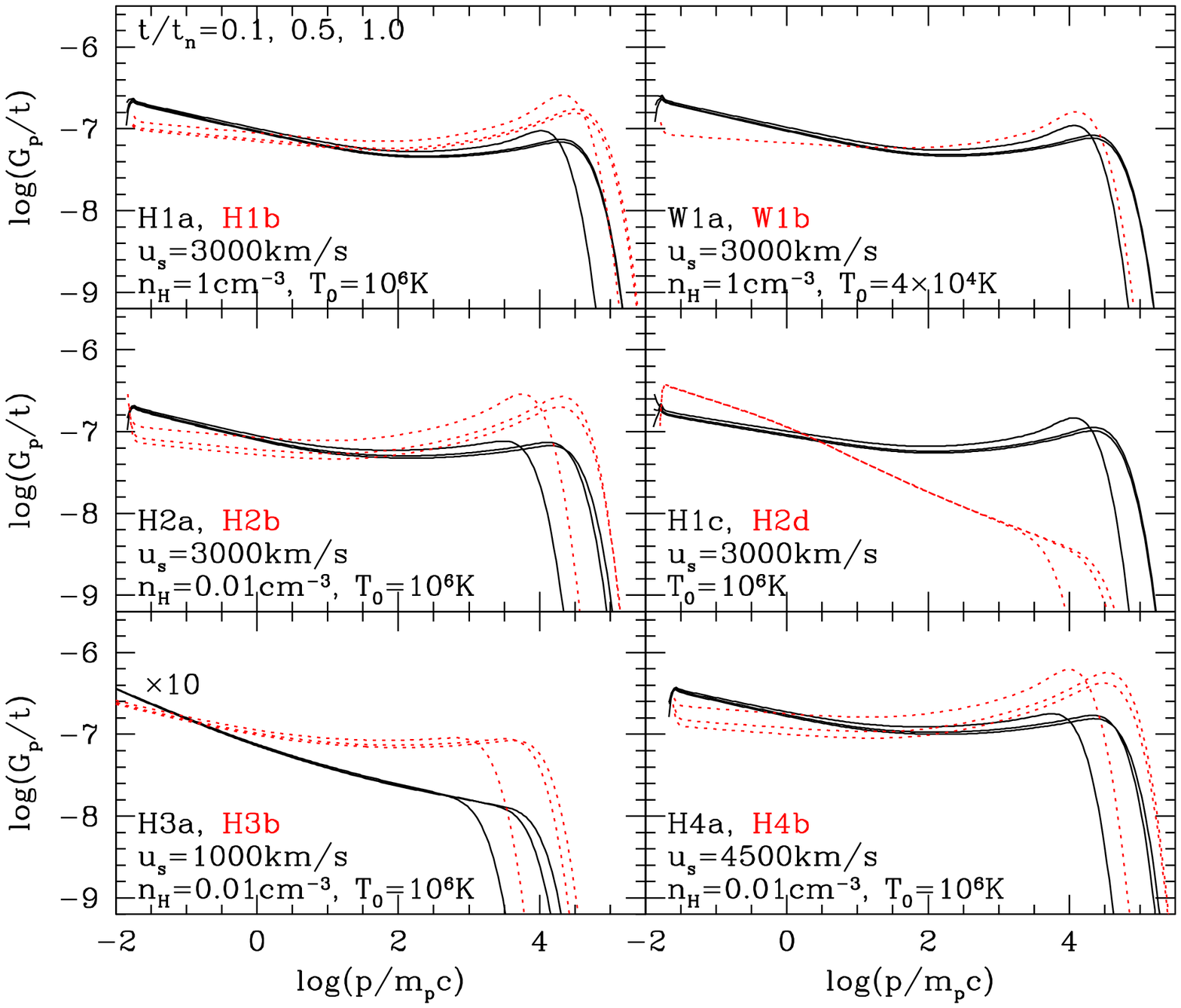}}
\vskip -1.5cm
\caption{
Volume integrated distribution function of CR protons for different models:
H1a, W1a, H2a, H1c, H3a, H4a, (black solid lines), H1b, W1b, H2b, H2d, H3b, H4b (red dotted lines).
For W1b model, only the curve for $t/t_n=0.1$ is shown, because the simulation was terminated afterward.
The curves for H3a/b are multiplied by a factor of 10 to show them in the same scale as other models.
See Table 1 for model parameters.
}
\label{fig5}
\end{figure*}

\subsection{Planar Shock Parameters}

We consider planar shocks with $u_s=1000-4500\kms$, propagating into a uniform ISM 
magnetized with $B_0=5\muG$.
The model parameters are summarized in Table 1.
Previous studies have shown that
the shock sonic Mach number is one of the key parameter governing the evolution 
and the DSA efficiency \citep[e.g.][]{kj07,kang09},
so here two phases of the ISM are considered:
the {\it warm phase} with $T_0=4\times 10^4$K (W models),
and the {\it hot phase} with $T_0=10^6$K (H model).
The sonic Mach number of each model is given as $M_s=20(T_0/10^6{\rm K})^{-1/2}u_{3000}$,
where $u_{3000}=u_s/3000\kms$.
Two values of the gas density $n_H=0.01 \cm3$ and $1 \cm3$ are considered.
The upstream Alfv\'en speed is then $v_{A,0}=B_0/\sqrt{4\pi\rho_0}= (18.3 \kms) n_H^{-1/2}$,
so the Alfv\'enic Mach number is  $M_{A,0}=u_s/v_{A,0}=164 \sqrt{n_H} u_{3000}$.

We consider W1a, H1a, H2a, H3a and H4a models as fiducial cases with canonical values of model parameters:
$f_A=1.0$ and $\omega_H=0.1$.
In models H1b, H2b, H3b and H4b, Alfv\'enic drift is turned off ($u_{w,1}=0$) for comparison.
But we note that these models are not self-consistent with our MFA model, which assumes that Alf\'ven waves
propagate along the amplified magnetic field.
In H1c model, MFA is reduced by setting $f_A=0.5$ and $\omega_H=0.5$ . 
Model H2d is chosen to see the effects of Alfv\'enic drift in the postshock region.

The physical quantities are normalized in the numerical code and in
the plots below by the following characteristic values:
$u_n = u_s $,
$x_n= R_s=3 {\rm~ pc}$,
$t_n= x_n/u_n = (978 {\yrs}) u_{3000}$,
$\kappa_n= u_n x_n$,
$\rho_n= (2.34\times 10^{-24} {\rm g ~cm^{-3}})\cdot n_H$,
and $P_n= \rho_n u_n^2 = (2.11\times 10^{-7} {\rm erg ~cm^{-3}})\cdot n_H u_{3000}^2$.

\begin{figure*}[t]
\vskip -1.0cm
\centerline{\epsfysize=14cm\epsfbox{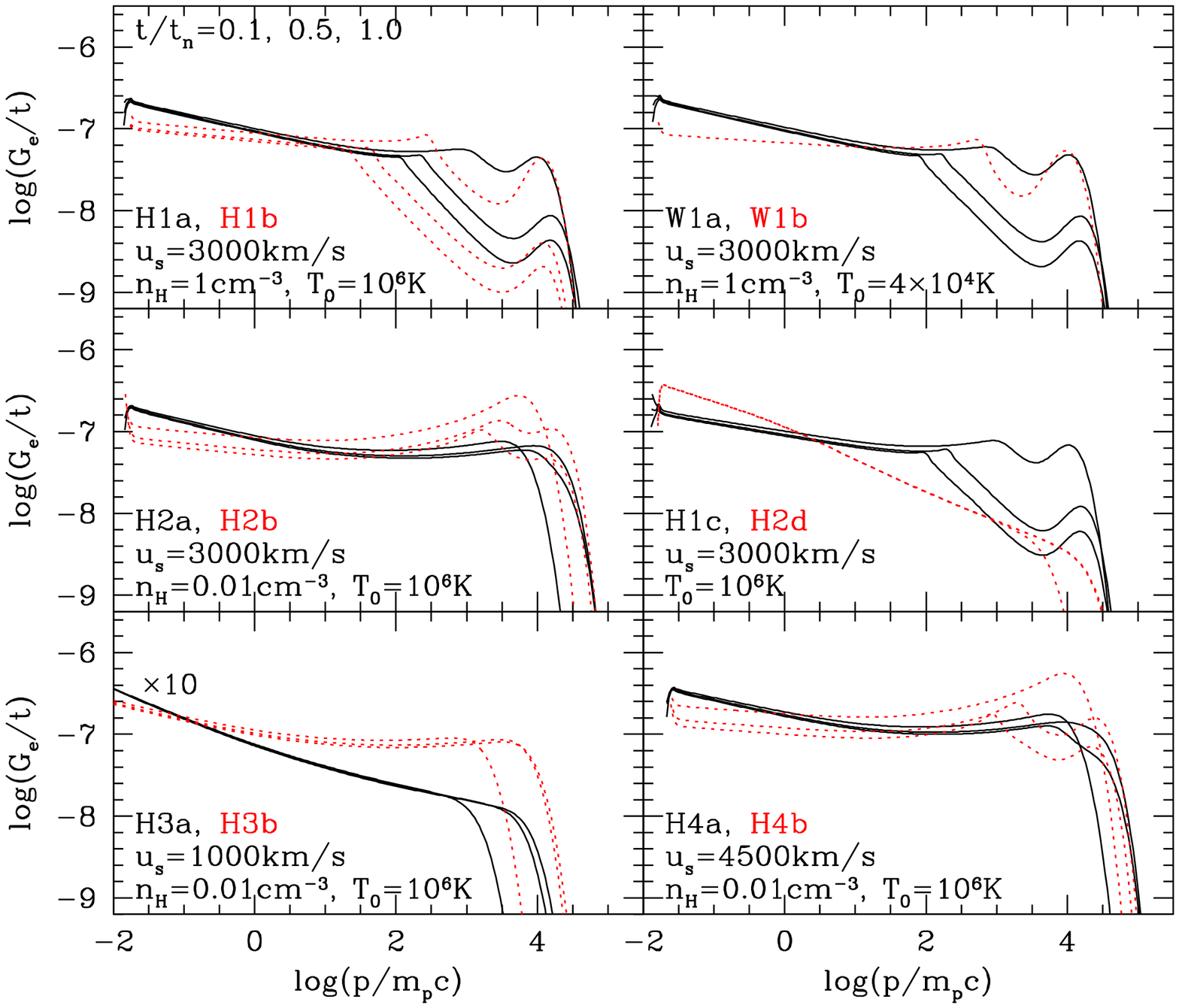}}
\vskip -1.5cm
\caption{
Same as Figure 5 except that the volume integrated distribution function of CR electrons are shown.
}
\label{fig6}
\end{figure*}

\section{DSA SIMULATION RESULTS}

Figures 1-3 show the spatial profiles of the model magnetic field, CR pressure, gas density, and the volume integrated distribution functions of protons ($G_p= \int g_p(x,p) dx$) and electrons ($G_e= \int g_e(x,p) dx$) for H1a and H1b ($M_{A,0}=164$), H2a and H2b ($M_{A,0}=16.4$),
and H3a and Hb ($M_{A,0}=5.46$) models, respectively. 
In these simulations, the highest level of refinement is $l_{\rm g,max}=8$ and the factor of each refinement is two, so the
ratio of the finest grid spacing to the base grid spacing is $\Delta x_8/\Delta x_0 = 1/256$ \citep{kjg02}.
Since Figures 1-3 show the flow structure on the base grid, the precursor profile may appear to be resolved rather poorly here.
More accurate values of the precursor density compression ($\rho_1$) and magnetic field amplification ($B_1$) can be 
found in Figure 4 below.
Also note that the FEB is located at $x_{\rm FEB}=0.3$pc and we use $K_{e/p}=0.1$ here
in order to show the proton and electron spectra together.

These figures demonstrate that 
1) the shock structure reaches the time-asymptotic state 
and evolves in a self-similar fashion for $t/t_n\ga 0.1$ \citep{kj07}.
2) Also the proton spectrum approaches the steady state for $t/t_n \ga 0.5$ due to the FEB,
while the electron spectrum continues to cool down in the downstream region.
3) Magnetic field is amplified by a greater factor for a higher $M_{A,0}$.
4) Alfv\'enic drift steepens the CR spectrum, and reduces the CR acceleration efficiency and
flow modification by CR feedback, resulting in lesser MFA.
5) At low energies the CR spectra are much steeper than the test-particle power-law due to the velocity
profile and magnetic field structure in the precursor.

In H1a model with Alfv\'enic drift (solid lines), 
the gas density immediately upstream of the subshock increases to
$\rho_1/\rho_0 \approx 1.2$, while the total compression ratio becomes $\rho_2/\rho_0 \approx 4.5$.
So the flow structure is moderately modified by the CR pressure feedback:
$U_1\approx 0.9$ and $P_{c,2}/\rho_0 u_s^2 \approx 0.12$.
Since the Alfv\'en Mach number is high ($M_{A,0}=164$), the self-amplified magnetic field strength, based on 
the model in equation (\ref{Bpre}), 
increases to $B_1\approx 100 \mu$G, which results in the immediate postshock fields of $B_2\approx 350 \mu$G.
Compared to the model without Alfv\'enic drift (H1b, dotted lines), the CR distribution functions are softer.
Although Alfv\'enic drift steepens the distribution function in H1a model,
$G_p(p)$ still exhibits a significant concave curvature and it is slightly 
flatter than the test-particle power-law ($E^{-2}$) at the highest energy end.
This is because $M_{A,0}=164$ is too large to induce the significant enough softening 
for $p\sim p_{\rm p, max}$ (see Equation (\ref{qt})). 
Note that $p_{\rm p, max}$ is lower in H1a model than that in H1b model, because of weaker MFA.

The structures of the integrated electron spectra are complex for $p \la p_{\rm e,max}$.
As shown in Equation (\ref{pbr}), the volume integrated electron energy spectrum steepens
by one power of the momentum due to radiative cooling.
One can see that the break momentum, $p_{\rm br}(t)$, shifts to lower momenta in time.
The peak near $p_{\rm e,max}$ comes from the electron population in the upstream region, which cools
much less efficiently due to weaker magnetic field there \citep{ekjm11}.
Since MFA is much stronger in H1b model, compared to H1a model,
the electron spectrum cools down to lower momentum in the downstream region.

Comparing H2a/b in Figure 2 with H1a/b in Figure 2, one can see that the degree of shock modification is similar
in the these models.
Because of a lower gas density ($n_H=0.01\cm3$) in H2a/b models, the Alfv\'en Mach number 
is smaller ($M_{A,0}=16.4$) and so MFA is much less efficient, compared to H1a/b models. 
In H2a model the amplified preshock field increases to only $B_1\approx 10\muG$, while 
the postshock field reaches $B_2\approx 35 \muG$.
Because of much weaker magnetic field, compared to H1a/b model, 
the electron spectra are affected much less by radiative cooling.

Since H3a/b models in Figure 3 have a lower sonic Mach Number ($M_s=6.7$),
the flow structures are almost test-particle like with $B_1\approx B_0$,
$\rho_2/\rho_0\approx 4$, and $P_{c,2}/\rho_0 u_s^2 \approx0.05-0.13$.
So the CR acceleration, flow modification, and MFA are all less efficient, 
compared to H1a/b and H2a/b models.
In H3a model, especially, the CR spectra are as steep as $E^{-2.1}-E^{-2.3}$
and electrons do not suffer significant cooling.

Figure 4 shows how various shock properties change in time for different models: 
the CR injection fraction, postshock CR pressure,
density compression factors and magnetic field strengths. 
As discussed above, the magnetic field amplification is more efficient in the models with higher $M_{A,0}$:
$B_1/B_0 \approx 20$ for $M_{A,0}=164$ (H1a, H1c, W1a models),
$B_1/B_0 \approx 3$ for $M_{A,0}=24.6$ (H4a),
$B_1/B_0 \approx 2$ for $M_{A,0}=16.4$ (H2a),
$B_1/B_0 \approx 1$ for $M_{A,0}=5.46$ (H3a).

According to previous studies, nonlinear DSA without self-consistent MFA and Alfv\'enic drift
predicts that the DSA efficiency depends strongly on the sonic Mach number $M_s$ and the
CR pressure asymptotes to $P_{c,2}/\rho_0u_s^2 \sim 0.5$ for $M_s\ga 20$.
\citep{kj07,kang09}.
However, Figure 4 shows that in the models with MFA and Alfv\'enic drift  
the CR acceleration and MFA are reduced in such a manner that the DSA efficiency
saturates roughly at $P_{c,2}/\rho_0u_s^2 \sim 0.1$ for $20\la M_s\la 100$.
We can see that models with a wide range of sonic Mach number,
i.e. W1a($M_s=100$), H4a ($M_s=30$), H2a and H1a ($M_s=20$), all
have similar results: $\rho_2/\rho_0\approx 4.5$, and
$P_{c,2}/\rho_0 u_s^2 \approx 0.1$.

Figures 5 and 6 show the volume-integrated
distribution function, $G_p(p)/(n_0 u_s t)$, for protons and, $G_e(p)/(n_0 u_s t)$, 
for electrons, respectively, for different models.
Again the proton spectrum approaches to the steady state for $t/t_n \ga 0.5$, 
when $p_{\rm p,max}(t)$ satisfies the condition, $\kappa(p_{\rm p,max})/u_s \sim x_{\rm FEB}$.
We note that for W1b model only the curve at $t/t_n=0.1$ is shown, because the simulation was
terminated when the subshock disappears because of very efficient DSA.
These figures demonstrate that the CR spectra is steepened by Alfv\'enic drift, especially at lower energies,
and that the degree of softening is greater for smaller $M_{A,0}$.
In H2d model, in which the downstream drift is included ($u_{w,2}=-v_A$) in addition to the upstream drift, 
the CR spectra are steepened drastically.

In the volume-integrated electron spectrum, the low-energy break corresponds to the momentum at
which the electronic synchrotron/IC loss time equals the shock age.
In the models with stronger magnetic field (e.g., H1a and W1a models),
this spectral break occurs at a lower $p_{\rm e,br}$, and 
the separate peak around $p_{\rm e,max}$ composed of the upstream population
becomes more prominent.

\section{SUMMARY}

Using the kinetic simulations of diffusive shock acceleration at planar shocks,
we have calculated the time-dependent evolution of the CR proton and electron spectra
for the shock parameters relevant for typical young supernova remnants.
In order to explore how various wave-particle interactions affect the DSA process, 
we adopted the following phenomenological models: 
1) magnetic field amplification (MFA) induced by CR streaming instability in the precursor,
2) drift of scattering centers with Alfv\'en speed in the amplified magnetic field, 
3) particle injection at the subshock via thermal leakage injection, 
4) Bohm-like diffusion coefficient, 
5) wave dissipation and heating of the gas,
and 6) escape of highest energy particles through a free escape boundary.

The MFA model assumes that the amplified magnetic field is isotropized by a variety of turbulent 
processes and so the Alfv\'en speed is determined by the
{\it local amplified} magnetic field rather than the background field \citep{capri12}.
This model predicts the magnetic field amplification factor scales with the upstream Alfv\'enic Mach number
as $B_1/B_0 \propto M_{A,0}$, and also increases with the strength of the shock precursor (see Equation (\ref{Bpre})).

Moreover, we assume that self-generated MHD waves drift away from the shock with respect to the background
flow, leading to smaller velocity jumps that particles experience scattering across the shock.
The ensuing CR distribution function becomes steeper than that calculated without Alfv\'enic drift,
so the CR injection/acceleration efficiencies and the flow modification due to CR feed back are reduced.

The expected power-law slope depends on the Alfv\'enic Mach number as given in Equations (\ref{qs})-(\ref{qt}). 
With our MFA model that depends on the precursor modification, 
the upstream Alfv\'enic drift affects lower energy particles more strongly,
steepening the low energy end of the spectrum more than the high energy end.
Hence, for $M_{A,0}\ga 10$, the CR spectra still retain the concave curvature
and they can be slightly flatter than $E^{-2}$ at the high energy end.
For weaker shocks with $M_s=6.7$ and $M_{A,0}=5.5$ (H3a model), on the other hand,
the Alfv\'enic drift effects are more substantial, so
the energy spectrum becomes as steep as $N(E)\propto E^{-2.1}-E^{-2.3}$.

We can explain how MFA and Alfv\'enic drift regulate the DSA as follows.
As CR particles stream upstream of the shock, magnetic field is amplified and Alfv\'en speed in the local
$B(x)$ increases in the precursor.
Then scattering centers drift with enhanced $v_A$, the CR spectrum is steepened and the CR acceleration
efficiency is reduced, which in turn restrict the growth of the precursor \cite[see also][]{capri12}. 
So the flow modification due to the CR pressure is only moderate with $\rho_2/\rho_0 \approx 4.5$.
As a result, the DSA efficiency saturates roughly at $P_{c,2}/\rho_0 u_s^2 \sim 0.1$ for $20\la M_s\la 100$.
For $M_s=20$ shocks with $u_s=3000\kms$, for example, 
in the models with Alfv\'enic drift (H1a and H2a),
the CR injection fraction is reduced from $\xi \sim 2\times 10^{-3}$  
to $\sim 2\times 10^{-4}$,  
while the CR pressure decreases from $P_{c,2}/\rho_0u_s^2 \sim 0.25$ to $ \sim 0.12$,
compared to the model without Alfv\'enic drift (H1b and H2b)
(see Figure 4).

This study demonstrates that detailed nonlinear treatments of wave-particle interactions 
govern the CR injection/acceleration efficiencies 
and the spectra of CR protons and electrons. Thus it is crucial to understand in a quantitative way 
how plasma interactions amplify magnetic field and
affect the transportation of waves in the shock precursor through detailed plasma simulations such as
PIC and hybrid simulations.
Moreover, the time-dependent behaviors of self-amplified magnetic field and CR injection as
well as particle escape will determine the spectra of the highest energy particles accelerated at astrophysical shocks.
We will present elsewhere the results from more comprehensive DSA simulations 
for a wide range of sonic and Alfv\'en Mach numbers.

\acknowledgments{
This research was supported by Basic Science Research Program through
the National Research Foundation of Korea (NRF) funded by the Ministry
of Education, Science and Technology (2012-001065).
}


\end{document}